# CSTNet: A Dual-Branch Convolutional Network for Imaging of Reactive Flows using Chemical Species Tomography

Yunfan Jiang, Jingjing Si, Rui Zhang, Godwin Enemali, *Member, IEEE*, Bin Zhou, Hugh McCann, Chang Liu, *Member, IEEE*

*Abstract*—Chemical Species Tomography (CST) has been widely used for *in situ* imaging of critical parameters, e.g. species concentration and temperature, in reactive flows. However, even with state-of-the-art computational algorithms the method is limited due to the inherently ill-posed and rank-deficient tomographic data inversion, and by high computational cost. These issues hinder its application for real-time flow diagnosis. To address them, we present here a novel CST-based convolutional neural Network (CSTNet) for high-fidelity, rapid, and simultaneous imaging of species concentration and temperature. CSTNet introduces a shared feature extractor that incorporates the CST measurement and sensor layout into the learning network. In addition, a dual-branch architecture is proposed for image reconstruction with crosstalk decoders that automatically learn the naturally correlated distributions of species concentration and temperature. The proposed CSTNet is validated both with simulated datasets, and with measured data from real flames in experiments using an industry-oriented sensor. Superior performance is found relative to previous approaches, in terms of robustness to measurement noise and millisecond-level computing time. This is the first time, to the best of our knowledge, that a deep learning-based algorithm for CST has been experimentally validated for simultaneous imaging of multiple critical parameters in reactive flows using a low-complexity optical sensor with severely limited number of laser beams.

## I. Introduction

IN the past two decades, Chemical Species Tomography (CST) has been widely applied for non-intrusive and sensitive imaging of multiple critical flow parameters, e.g. gas-phase species concentration [1-4], temperature [3-4], and velocity [5]. To solve the inverse problem of CST, a variety of computational tomographic algorithms have been developed, some adapted from previous hard-field tomography modalities, and improved subject to the characteristics of the flow field. These algorithms can be mainly categorised as:

- Algebraic techniques based on linear back projection, e.g. algebraic reconstruction technique (ART) [6] and Landweber algorithm [2], [7];
- Regularisation methods, e.g. Tikhonov regularisation [8];
- Global optimisation, e.g. simulated annealing [9];
- Statistical inversion, e.g. covariance estimation [10]; and
- Dimensional reduction techniques, e.g. surrogate functions method [11].

As discussed in [12], these algorithms can, to some extent, mitigate the difficulty of robust image reconstruction in CST. Nevertheless, deficiencies still remain, particularly the appearance of artefacts in the reconstructed images due to the rank-deficient tomographic data, and the high computational cost incurred due to the complex, mostly iteration-based, mathematical operations. These issues severely hinder the exploitation of CST for applications requiring high-fidelity performance and real-time image reconstruction.

The last 10 years have witnessed a boom in the use of deep learning for bioimaging and medical imaging [13-15]. As one of the most prevalent deep learning architectures, Convolutional Neural Network (CNN) [16] is a good candidate to overcome the above-mentioned issues in CST due to the following properties:

1) Automatic discovery of intricate features. In comparison with the computational CST algorithms with constraints manually imposed on the flow fields, CNN can automatically construct data representations during the learning process [17], enabling end-to-end (i.e. from measurements to reconstructed images) learning of intricate features of the flow fields with superior generalisation ability.
2) Accurate image retrieval with strong robustness. CNN can capture and learn distinct features of the flow fields without overfitting the tomographic data, thus yielding strong immunity to measurement noise. As the training sets are established from CST measurements of noise-free target fields, CNN trained under supervision intrinsically addresses the limited sampling of CST systems, thus helping to eliminate artefacts in the tomographic images.

This work was supported in part by the UK Engineering and Physical Sciences Research Council under Grant EP/P001661/1. *(Corresponding author: Chang Liu.)*

Y. Jiang, J. Si, R. Zhang, G. Enemali, H. McCann and C. Liu are with the School of Engineering, University of Edinburgh, Edinburgh EH9 3JL, U.K. (e-mail: C.Liu@ed.ac.uk).

J. Si is also with the School of Information Engineering, Yanshan University, Qinhuangdao 066004, China.

B. Zhou is with the School of Energy and Environment, Southeast University, Nanjing 210018, China.

23) Rapid and cost-effective computation, empowered by hardware acceleration, for rapid processing of the large amount of CST data typically acquired. This has the potential to facilitate online imaging, and thus real-time flow and combustion diagnosis.

CNN has been demonstrated recently in CST simulations to perform spatially resolved measurements in combustion diagnosis [18], [19]. In addition, CNN has been applied in a proof-of-concept experiment to reconstruct the three-dimensional distribution of methane concentration using mid-infrared CST [20]. Although these endeavours are promising for industrial application of CNN in CST, the following three issues remain to be addressed as a matter of urgency:

A. The properties of the CST measurement system. The only *a priori* information taken into account by the previous work was that pertaining to the attributes of the phantoms themselves, i.e. smooth distributions of species concentration and temperature. However, more in-depth features, e.g. smoothness and centrosymmetry in the CST measurement and sensor layout, were not considered in the learning architectures.

B. Inter-dependence of species concentration and temperature distributions. The previous work assumed independence between species concentration and temperature distributions, neglecting their internal correlation in combustion processes. Furthermore, these efforts were incapable of multi-parameter retrieval. They can only reconstruct either species concentration distribution or temperature distribution with a single network.

C. Practicality for industrial applications. Previous optical sensors used up to 6 angular views and tens of laser beams per view, greatly assisting the image reconstruction process. However, severely limited optical access, with fewer laser beams, is typical for industrial applications, e.g. for reliability maintenance. Such limited projection data place more rigorous challenges on the applicability of CNN in CST.

To address these issues, we propose here a novel network, named as CST-based convolutional neural Network (CSTNet). The remainder of this paper is organised as follows. Based on formulation of the inverse problem of CST, we first introduce the hybrid *a priori* information and the architecture of CSTNet in Section II. Then, the CSTNet model is established using the 32-beam CST sensor and adapting our specific task in Section III. Subsequently, we train the established neural network and examine its performance with a simulated test set in Section IV, and experiments in Section V, respectively. Finally, conclusions are presented in Section VI.

## II. METHODOLOGY

### A. Mathematical Formulation of CST

When a laser beam at frequency $v$ [cm$^{-1}$] penetrates an absorbing gas sample with a path of length $L$ [cm], the wavelength-dependent absorbance, $\alpha(v)$, is defined as

$$\alpha(v) = \ln \frac{I_0(v)}{I_t(v)} = \int_0^L P(l)X(l)S(T(l))\phi(v,l)dl, \quad (1)$$

where $I_0(v)$ and $I_t(v)$ are the incident and transmitted laser intensities, respectively. $l$ is the local position along the path, $P(l)$ [atm] the local pressure, $X(l)$ the local molar fraction of the absorbing species, $T(l)$ [K] the local temperature, $S(\cdot)$ [cm$^{-2}$atm$^{-1}$] the temperature-dependent line strength, and $\phi(\cdot)$ [cm] the line-shape function [12].

Since the line-shape function is normalised to unity, i.e. $\int_{-\infty}^{+\infty}\phi(v,l)dv \equiv 1$, the path integrated absorbance, $A_v$, can be formulated as

$$A_v = \int_{-\infty}^{+\infty}\alpha(v)dv = \int_0^L P(l)X(l)S(T(l))dl = \int_0^L a_v(l)dl, \quad (2)$$

where $a_v(l)$ is the local density of $A_v$.

The inverse problem of CST is formulated by discretising the Region of Interest (RoI) into $N$ pixels, as shown in Fig. 1. As a result, (2) is discretised as

$$\boldsymbol{A}_v = \boldsymbol{L}\boldsymbol{a}_v, \quad (3)$$

where $\boldsymbol{A}_v \in \mathbb{R}^{M\times 1}$ denotes the vector of path integrated absorbance obtained from $M$ CST measurements, with its element $A_{v,i}$ representing the path integrated absorbance of the $i$-th beam. $\boldsymbol{L} \in \mathbb{R}^{M\times N}$ is the sensitivity matrix with its element $L_{i,j}$ representing the length of the laser path segment for the $i$-th laser beam passing through the $j$-th pixel. $i \in \{1,2,\ldots,M\}$ and $j \in \{1,2,\ldots,N\}$ are the indices of laser beams and pixels, respectively. $\boldsymbol{a}_v \in \mathbb{R}^{N\times 1}$ is the vector of absorbance density with its elements $a_{v,j} = P_j X_j S(T_j)$.

### B. Hybrid a Priori Information

#### 1) Smoothness

The optical layout of the CST sensor is given by $Q$ angular views and $R$ parallel laser beams per view, satisfying $Q \times R = M$. Imposed by the adjacent arrangement of laser beams, the path integrated absorbance within the $q$-th angular view $\theta_q$, $\boldsymbol{A}_v^{\theta_q} \in \mathbb{R}^{R\times 1}$, experiences smooth change from beam to beam due to the smooth distributions of species concentration and temperature in the RoI:

$$\left|A_{v,r}^{\theta_q} - A_{v,r-1}^{\theta_q}\right| \leq \epsilon, \forall r \in \{2,3,\ldots,R\}, \forall q \in \{1,2,\ldots,Q\}, \quad (4)$$

where $\epsilon$ is a threshold and $A_{v,r}^{\theta_q}$ the $r$-th element of $\boldsymbol{A}_v^{\theta_q}$. Learning such *a priori* information regarding smoothness enables speedy convergence with a lower loss during training. Hence, it is incorporated into CSTNet detailed in Section II C.

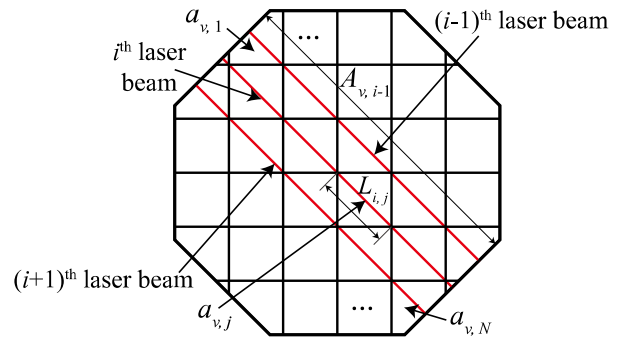

Fig. 1. Geometric description of CST measurements in the discretised RoI.

### 2) Centrosymmetry

CNN exploits the hierarchical property of images and therefore is superior for recognising and detecting patterns in the images [17]. This characteristic can assist to perceive the physical fields with CST. When the CST beam array is rotationally symmetric around the centre of the RoI, e.g. invariant on 180-degree rotation, the measurement patterns thus created can be exploited by CNN. We illustrate this property by the following example. Figs. 2 (a) and (b) show two "phantom" temperature distributions that are centrosymmetric with respect to each other with identical distributions of temperature except that they are rotated by 180 degrees around the centre of the RoI. Using the 32-beam CST sensor demonstrated later in Section III, path integrated absorbance at frequencies $v_1$ and $v_2$, i.e. $A_{v_1} \in \mathbb{R}^{32\times 1}$ and $A_{v_2} \in \mathbb{R}^{32\times 1}$, can be measured to carry out two-line temperature imaging [12]. The centrosymmetry, introduced by the CST measurement, can be characterised by a pattern, named the centrosymmetry heatmap, which contains information of both $A_{v_1}$ and $A_{v_2}$. Firstly, $A_{v_1}$ and $A_{v_2}$ are reshaped to $A_{v_1}^{reshape} \in \mathbb{R}^{4\times 8}$ and $A_{v_2}^{reshape} \in \mathbb{R}^{4\times 8}$, respectively. Then, the heatmap is constructed by concatenating $A_{v_1}^{reshape}$ at the upper half and vertically flipped $A_{v_2}^{reshape}$ at the lower half. As shown in Figs. 2 (c) and (d), the two phantom temperature distributions lead to heatmaps with identical patterns but opposite orientations. In other words, when a fixed centrosymmetric beam geometry is employed, moving the inhomogeneity to a centrosymmetric location in the RoI is equivalent to re-orientating the heatmap. Learning these heatmaps enables prediction of both the distributions of temperature values and the locations of inhomogeneities in the RoI. Therefore, centrosymmetry is learnt as *a priori* information in CSTNet.

### C. CSTNet Architecture

As shown in Fig. 3, the proposed CSTNet consists of two main parts, i.e. a shared feature extractor and a dual-branch architecture with crosstalk decoders. A Lambda layer is firstly used to generate two heatmaps from the projection data of CST. The two heatmaps are then directly learnt by the shared feature extractor [21]. The upper and lower branches of the feature extractor learn the centrosymmetry and smoothness, respectively. Both learning processes can be formulated by

$$O = f(W * I + b), \quad (5)$$

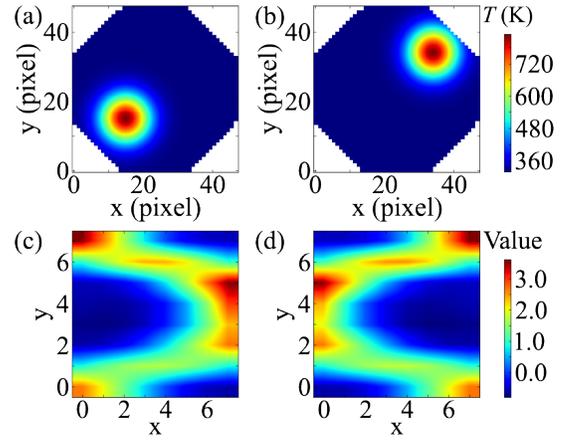

Fig. 2. Illustration of *a priori* information of centrosymmetry in CST. (a) and (b) are two centrosymmetric temperature images; (c) and (d) are the heatmaps containing $A_{v_1}$ and $A_{v_2}$ obtained from the temperature images in (a) and (b), respectively.

where $I \in \mathbb{R}^{H_I \times W_I \times C_I}$ is the input heatmaps or intermediate feature maps, $W \in \mathbb{R}^{H_W \times W_W \times C_W}$ the convolution kernel, $O \in \mathbb{R}^{H_O \times W_O \times C_O}$ the output feature maps, $b \in \mathbb{R}^{C_b \times 1}$ the bias vector, $f(\cdot)$ the activation function, and $*$ the operand for 2D convolution. $H_I(H_O)$, $W_I(W_O)$, and $C_I(C_O)$ are height, width, and channel of $I$ ($O$), respectively. $H_W$, $W_W$, and $C_W$ are the filter height, filter width, and number of filters of $W$, respectively. $C_b$ is the length of $b$.

Then, the output feature maps are concatenated and fed into the dual-branch architecture for simultaneous imaging of species concentration and temperature. In reactive flows, for example, hydrocarbon combustion processes, the species concentration distribution is generally correlated with the temperature distribution. Therefore, their correlation is incorporated into the dual-branch architecture [22] with crosstalk decoders. Both decoders consist of $G$ stages and can be simultaneously computed by

$$\mathcal{X}_g = f\left(h(\mathcal{X}_{g-1}) + \mathcal{W}_g^T \odot h(\mathcal{T}_{g-1})\right), g \in \{1, 2, ..., G\} \quad (6)$$

and

$$\mathcal{T}_g = f\left(h(\mathcal{T}_{g-1}) + \mathcal{W}_g^{\mathcal{X}} \odot h(\mathcal{X}_{g-1})\right), g \in \{1, 2, ..., G\}, \quad (7)$$

where $h(\cdot)$ denotes operation before the activation function $f(\cdot)$. $\mathcal{X}_g \in \mathbb{R}^{N_g \times 1}$ and $\mathcal{T}_g \in \mathbb{R}^{N_g \times 1}$ are the outputs from the $g$-th stage of concentration and temperature decoders, respectively. $N_g$ represents the length of the output from the $g$-th stage. $\mathcal{W}_g^T \in$

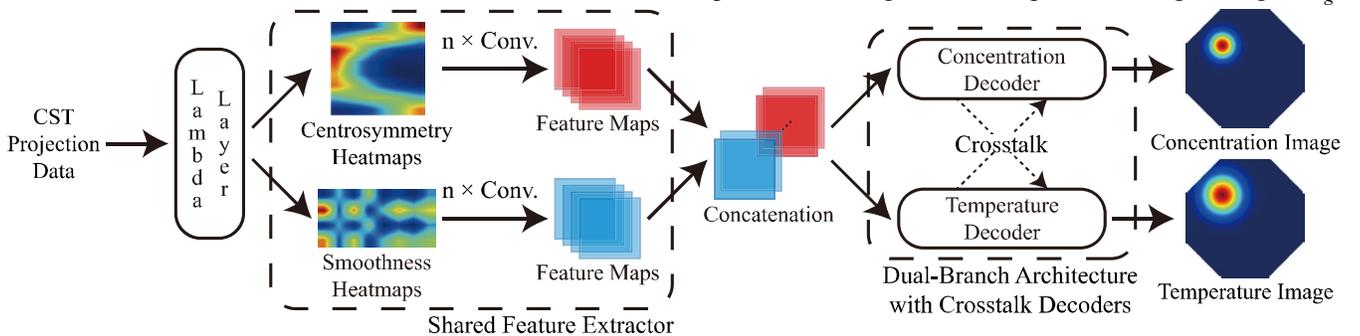

Fig. 3. The overall architecture of CSTNet.

$\mathbb{R}^{N_g \times 1}$ and $\mathcal{W}_g^{\mathcal{X}} \in \mathbb{R}^{N_g \times 1}$ are the crosstalk weight vectors in the $g$-th stage of the concentration and temperature decoders, respectively. $\odot$ denotes the element-wise production.

The selections of $h(\cdot)$ and $f(\cdot)$ are subject to different tasks and purposes. For instance, $h(\cdot)$ can be either fully-connected (FC), deconvolutional [23], FC with batch normalisation [24], or deconvolutional with batch normalisation. $f(\cdot)$ can be either ReLU [25], Leaky ReLU [26], or PReLU [27]. In Section III, we will establish our task-specific model and detail the implementation of CSTNet.

## III. MODEL ESTABLISHMENT

### A. System Specification

A CST sensor with 32 beams, as shown in Fig. 4, is used in this work to generate 4 equiangular projections, i.e. $Q = 4$, each with 8 equispaced parallel laser beams, i.e. $R = 8$ [28]. The angular spacing between projections is 45°, and neighbouring beams within each projection are separated by 1.80 cm. The distance between an emitter and a receiver is 36.76 cm. The RoI is defined as the octagonal sensing area with the side length of 12.60 cm. The dimension of each pixel in the RoI is 0.766 cm × 0.766 cm, resulting in 1924 uniformly segmented pixels, i.e. $N = 1924$.

As a principal product of hydrocarbon combustion, water vapour ($H_2O$) has a strong near-infrared absorption spectrum and therefore is selected as the target absorption species in this work. Two $H_2O$ transitions at $v_1 = 7185.6$ cm$^{-1}$ and $v_2 = 7444.36$ cm$^{-1}$ are adopted given their good temperature sensitivity for the target temperature range of 300-1500 K [4]. Using the 32-beam CST sensor the projection data, $A_{v_1} \in \mathbb{R}^{32 \times 1}$ and $A_{v_2} \in \mathbb{R}^{32 \times 1}$, are obtained at $v_1$ and $v_2$, respectively. The end-to-end CSTNet will be implemented in the following subsection to simultaneously reconstruct the distributions of $H_2O$ concentration and temperature using $A_{v_1}$ and $A_{v_2}$.

### B. Implementation of CSTNet

#### 1) Lambda Layer

The Lambda layer is used to construct the centrosymmetry heatmap, $\mathfrak{S} \in \mathbb{R}^{8 \times 8 \times 1}$, and the smoothness heatmap, $\mathfrak{P} \in \mathbb{R}^{4 \times 8 \times 2}$, by combining and rearranging $A_{v_1}$ and $A_{v_2}$ according to Algorithm 1. With the Lambda layer, $A_{v_1}$ and $A_{v_2}$ can be directly used as inputs for the shared feature extractor, facilitating the end-to-end learning.

---

**Algorithm 1** Construction of centrosymmetry heatmaps and smoothness heatmaps from CST projection data

**Input:** Number of equiangular projections $Q$, number of equispaced parallel beams $R$, and CST measurements $A_{v_1} \in \mathbb{R}^{M \times 1}$ and $A_{v_2} \in \mathbb{R}^{M \times 1}$ where $M = Q \times R$

**Output:** Centrosymmetry heatmap $\mathfrak{S} \in \mathbb{R}^{2Q \times R \times 1}$ and smoothness heatmap $\mathfrak{P} \in \mathbb{R}^{R/F \times QF \times 2}$ with $F$ the least prime factor greater than 1 of $R$

**Initialise:** Empty $\mathfrak{S}$ and $\mathfrak{P}$

1:  $F \leftarrow$ the least prime factor greater than 1 of $R$
2:  Construct $\mathfrak{S}$:
     for $i$ in range(2) do
3:     $A_{v_i}^{reshape} \leftarrow$ reshape($A_{v_i}$, $(Q, R)$)
4:     if $v_i$ is the counterpart frequency then
5:      ($Q \times i$)-th to ($Q \times (i+1)-1$)-th rows of $\mathfrak{S} \leftarrow$ vertical_flip($A_{v_i}^{reshape}$)
6:     else
7:      ($Q \times i$)-th to ($Q \times (i+1)-1$)-th rows of $\mathfrak{S} \leftarrow A_{v_i}^{reshape}$
8:     end if
9:  end for
10: Construct $\mathfrak{P}$:
     for $i$ in range(2) do
11:    for $j$ in range($Q$) do
12:     $A_{v_i}^{patch} \leftarrow$ reshape(($R \times j$)-th to ($R \times (j+1)-1$)-th elements of $A_{v_i}$, $(\frac{R}{F}, F)$)
13:     ($F \times j$)-th to ($F \times (j+1)-1$)-th columns of the $i$-th channel of $\mathfrak{P} \leftarrow A_{v_i}^{patch}$
14:    end for
15: end for

---

#### 2) Shared Feature Extractor

The shared feature extractor takes the output of the Lambda layer as inputs and directly learns centrosymmetry from $\mathfrak{S}$ and smoothness from $\mathfrak{P}$ by convolution. Its outputs are shared by two decoders.

As detailed in Table I, the first branch that learns the centrosymmetry consists of three convolutional blocks. Each block adopts PReLU as the activation function. Batch normalisation is employed for accelerated training and can largely prevent overfitting. There are two reasons of not adopting pooling layers in CSTNet. First, the use of pooling is detrimental to the pixelwise prediction that we aim to achieve [29]. Second, CSTNet can hardly benefit from the reduced dimensions of feature maps, which are already of small dimensions, by using pooling layers. The forward propagation is computed by

$$\mathcal{S}_i = \begin{cases} \mathfrak{S}, i = 0 \\ \text{PReLU}_i\left(\text{BN}_{\gamma_i, \beta_i}\left(\mathcal{W}_i * \mathcal{S}_{i-1}\right)\right), i \in \{1, 2, 3\}, \end{cases} \quad (8)$$

where $\mathcal{W}_i$ is the convolution kernel for the $i$-th convolution block, $\mathcal{S}_i$ the $i$-th intermediate feature maps extracted from $\mathfrak{S}$, PReLU$_i(\cdot)$ the PReLU activation function for the $i$-th block, and BN$_{\gamma_i, \beta_i}(\cdot)$ the batch normalisation in the $i$-th block.

The second branch that learns the smoothness contains a single convolution block, in which PReLU and batch normalisation are adopted as well. To correctly extract the smoothness information, filters in $W$ are specifically designed with the size of $2 \times 2$ and strides of $(1, 2)$ such that they always operate on projection data obtained from adjacent receivers within the same projection angle. The forward propagation is computed by

$$\mathcal{P} = \text{PReLU}\left(\text{BN}_{\gamma,\beta}\left(W * \mathfrak{P}\right)\right), \quad (9)$$

where $\mathcal{P}$ is the extracted feature maps from $\mathfrak{P}$.

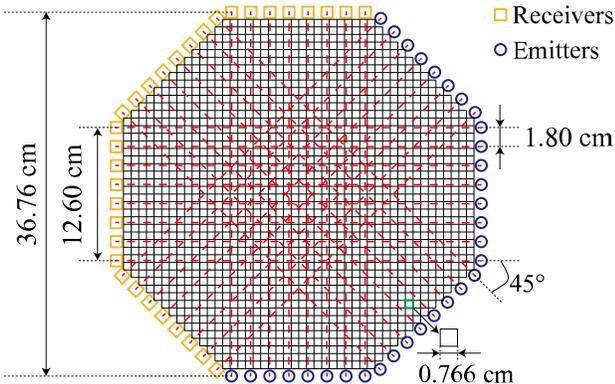

Fig. 4. Optical layout of the 32-beam CST sensor.

TABLE I. Detailed parameters for implementing CSTNet. The input dimension of each convolutional layer is described in the form of $H_I \times W_I \times C_I$. The filter sizes and strides are described in forms of $H_W \times W_W$ and (height, width). The input dimension of a certain layer is the output dimension of its previous layer.

| | Shared feature extractor | | | | Concatenation node | Dual-branch architecture with crosstalk decoders | | | | | |
|---|---|---|---|---|---|---|---|---|---|---|---|
| | Centro-symmetry extractor | Block 1 | Block 2 | Block 3 | | Conc. decoder | | Stage 1 | Stage 2 | Stage 3 | Stage 4 |
| Input dim. | | 8×8×1 | 8×8×64 | 8×8×128 | 6×6×256 | | Input dim. | 9984 | 8192 | 4096 | 2048 |
| Filter size | | 3×3 | 3×3 | 3×3 | | | | | | | |
| Stride | | (1, 1) | (1, 1) | (1, 1) | - | | | | | | |
| Padding | | 1 | 1 | 0 | | | | | | | |
| | Smoothness extractor | Block 1 | | | | Temp. decoder | | Stage 1 | Stage 2 | Stage 3 | Stage 4 |
| Input dim. | | 4×8×2 | | | 3×4×64 | | Input dim. | 9984 | 8192 | 4096 | 2048 |
| Filter size | | 2×2 | | | | | | | | | |
| Stride | | (1, 2) | | | - | | | | | | |
| Padding | | 0 | | | | | | | | | |

As a result, multi-scale feature maps, $\mathcal{S}_3 \in \mathbb{R}^{6 \times 6 \times 256}$ and $\mathcal{P} \in \mathbb{R}^{3 \times 4 \times 64}$, are extracted and further concatenated [30], yielding a vector containing latent features, $\mathcal{L} \in \mathbb{R}^{9984 \times 1}$.

*3) Dual-Branch Crosstalk Decoders*

$\mathcal{L}$ is fed into the dual-branch architecture with crosstalk decoders for simultaneous imaging of $H_2O$ concentration and temperature. In this case, FC layers are employed to fuse $\mathcal{L}$ [31]. The crosstalk decoders consist of 4 stages, with each stage depicted in Fig. 5.

With batch normalisation and PReLU activation, outputs from each of the first three stages can be computed through

$$\mathcal{X}_i = \begin{cases} \mathcal{L}, i = 0 \\ \text{PReLU}_i\left(\text{BN}_{\gamma_i^{\mathcal{X}},\beta_i^{\mathcal{X}}}\left(W_i^{\mathcal{X}} \mathcal{X}_{i-1}\right) + \mathcal{W}_i^{\mathcal{T}} \odot \text{BN}_{\gamma_i^{\mathcal{T}},\beta_i^{\mathcal{T}}}\left(W_i^{\mathcal{T}} \mathcal{T}_{i-1}\right)\right), i \in \{1,2,3\} \end{cases} \quad (10)$$

and

$$\mathcal{T}_i = \begin{cases} \mathcal{L}, i = 0 \\ \text{PReLU}_i\left(\text{BN}_{\gamma_i^{\mathcal{T}},\beta_i^{\mathcal{T}}}\left(W_i^{\mathcal{T}} \mathcal{T}_{i-1}\right) + \mathcal{W}_i^{\mathcal{X}} \odot \text{BN}_{\gamma_i^{\mathcal{X}},\beta_i^{\mathcal{X}}}\left(W_i^{\mathcal{X}} \mathcal{X}_{i-1}\right)\right), i \in \{1,2,3\} \end{cases}, \quad (11)$$

where $\text{BN}_{\gamma_i^{\mathcal{X}},\beta_i^{\mathcal{X}}}(\cdot)$ and $\text{BN}_{\gamma_i^{\mathcal{T}},\beta_i^{\mathcal{T}}}(\cdot)$ are the $i$-th stage batch normalisation in $H_2O$ concentration and temperature decoders.

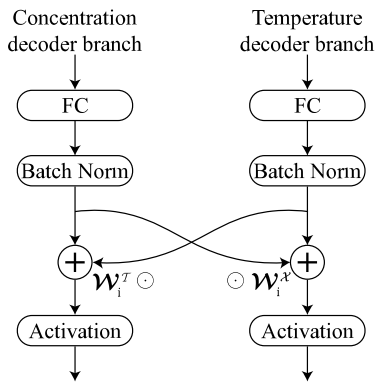

Fig. 5. A stage of crosstalk decoders.

The last stage, i.e. the output stage, is formed by imposing physical constraints on the $H_2O$ concentration and temperature. Hyperbolic Tangent function, $\text{Tanh}(\cdot)$, is adopted as the output activation considering that $H_2O$ concentration and temperature are supposed to be within a physically reasonable range. Therefore, the distributions of $H_2O$ concentration and temperature are finally reconstructed by

$$X = \text{Tanh}\left(\text{BN}_{\gamma_4^{\mathcal{X}},\beta_4^{\mathcal{X}}}\left(W_4^{\mathcal{X}} \mathcal{X}_3\right) + \mathcal{W}_4^{\mathcal{T}} \odot \text{BN}_{\gamma_4^{\mathcal{T}},\beta_4^{\mathcal{T}}}\left(W_4^{\mathcal{T}} \mathcal{T}_3\right)\right) \quad (12)$$

and

$$T = \text{Tanh}\left(\text{BN}_{\gamma_4^{\mathcal{T}},\beta_4^{\mathcal{T}}}\left(W_4^{\mathcal{T}} \mathcal{T}_3\right) + \mathcal{W}_4^{\mathcal{X}} \odot \text{BN}_{\gamma_4^{\mathcal{X}},\beta_4^{\mathcal{X}}}\left(W_4^{\mathcal{X}} \mathcal{X}_3\right)\right). \quad (13)$$

## IV. NETWORK TRAINING AND TESTING

### A. Dataset

The constructed dataset includes three categorises of two-dimensional (2D) distributions of $H_2O$ concentration and temperature with one, two, and three inhomogeneities. Each inhomogeneity is modelled by a 2D Gaussian profile. As noted in Section II C, $H_2O$ concentration distribution is generally well-correlated with temperature distribution. In each phantom, the peak locations of the inhomogeneities in $H_2O$ concentration distributions are modelled same as those in the temperature distributions. In general, spread of $H_2O$ concentration depends on flow convection, which is slower than heat transfer and dissipation. Therefore, the 2D Gaussian inhomogeneities in the $H_2O$ concentration distributions are generated with smaller standard deviations than those in the temperature distributions. To be specific, the distributions of $H_2O$ concentration and temperature are mathematically expressed as

$$X(x,y) = X_{min} + \sum_{d=1}^{D} \xi_d \left(X_{max} - X_{min}\right) \exp\left(-\frac{\left(x - x_c^d\right)^2 + \left(y - y_c^d\right)^2}{\sigma_X^2}\right) \quad (14)$$

and

$$T(x,y) = T_{min} + \sum_{d=1}^{D} \xi_d \left(T_{max} - T_{min}\right) \exp\left(-\frac{\left(x - x_c^d\right)^2 + \left(y - y_c^d\right)^2}{\sigma_T^2}\right), \quad (15)$$

where $x$ and $y$ denote the horizontal and vertical coordinates of the RoI, respectively. $(x_c^d, y_c^d)$ is the central position of the $d$-th Gaussian profile. $D$ is the total number of inhomogeneities in the phantom. $X_{max}$ ($T_{max}$) and $X_{min}$ ($T_{min}$) are the maximum and minimum $H_2O$ concentration (temperature), respectively. $\xi_d \sim$

$U(0.7, 1)$ is a random scaling factor. $\sigma_X$ and $\sigma_T$ are the standard deviations of $H_2O$ concentration and temperature inhomogeneities that satisfy $\sigma_X = \rho \sigma_T$ with $\rho \sim U(1/3, 1)$.

In this work, we adopted $X_{min} = 0.01$, $X_{max} = 0.12$, $T_{min} = 318$ K, and $T_{max} = 1300$ K. The dataset is generated with 19227 independent examples, which are then randomly divided into a training set with 13440 examples, a validation set with 5760 examples, and a test set with 27 examples. Using the $H_2O$ transitions at $v_1$ and $v_2$, 19227 sets of $A_{v_1}$ and $A_{v_2}$ are generated according to the forward problem formulated in (3).

Subsequently, path integrated absorbance for the $i$-th beam, $A_{v,i} \in \mathbb{R}^{E \times 1}$, in the training and validation sets are standardised to $(A_{v,i})^{processed} \in \mathbb{R}^{E \times 1}$ by

$$\left(A_{v,i}\right)^{processed} = \frac{A_{v,i} - \mu_{A_{v,i}}}{\sigma_{A_{v,i}}}, \quad (16)$$

where $E$ is the total number of examples in the training and validation sets. $\mu_{A_{v,i}}$ and $\sigma_{A_{v,i}}$ are the mean value and standard deviation of $A_{v,i}$ given by

$$\mu_{A_{v,i}} = \frac{1}{E}\sum_{e=1}^{E} A_{v,i}^e \quad (17)$$

and

$$\sigma_{A_{v,i}} = \sqrt{\frac{\sum_{e=1}^{E}\left(A_{v,i}^e - \mu_{A_{v,i}}\right)^2}{E}}, \quad (18)$$

with $A_{v,i}^e$ the $e$-th example of $A_{v,i}$.

The standardisation has two benefits. First, it can speed up the training process as the averages of input features are moved close to zero and their covariances are kept approximately same, which balances out the learning speed of weights connected to input nodes [32]. Second, standardisation can suppress generalisation error during image reconstruction, which is caused by bias and fluctuations in real CST measurements.

### B. Training Details

CSTNet is trained by minimising a weighted mean-square-error (MSE) loss in terms of the reconstructed and true distributions of $H_2O$ concentration and temperature. The trainable weights of CSTNet, $\psi$, is formulated by

$$\psi = \underset{\psi}{\operatorname{argmin}}\left\{\tau \frac{1}{N}\sum_{j=1}^{N}\left(T_j - \widehat{T}_j\right)^2 + (1-\tau)\frac{1}{N}\sum_{j=1}^{N}\left(X_j - \widehat{X}_j\right)^2\right\}, \quad (19)$$

where $\tau$ is a hyperparameter to trade off the MSE loss on imaging of $H_2O$ concentration and temperature. $\widehat{X}_j$ ($\widehat{T}_j$) and $X_j$ ($T_j$) are the reconstructed and true $H_2O$ concentration (temperature) in the $j$-th pixel, respectively.

Adam optimiser [33] was employed with a learning rate of $10^{-3}$ found by range test [34] and default values for other hyperparameters. To avoid overfitting, L2 weight decay was also imposed with a penalty factor of $2 \times 10^{-6}$ determined by Monte Carlo estimation [35]. $\tau$ was set to 0.5 to remain equal importance of $H_2O$ concentration and temperature imaging. The model was trained for 350 epochs until it converged using a single NVIDIA Tesla P100 GPU on a PCIe-based server.

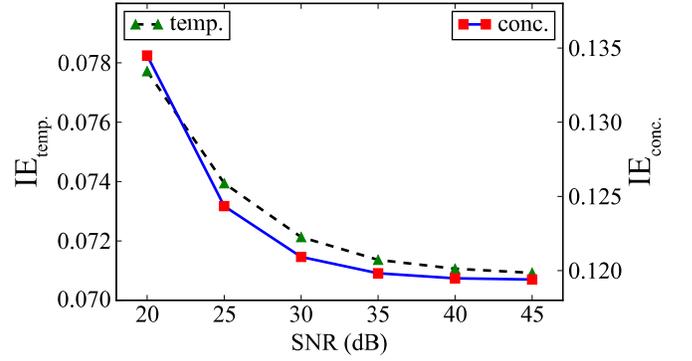

Fig. 6. Dependency of IEs for the reconstructed distributions of $H_2O$ concentration and temperature at different SNRs.

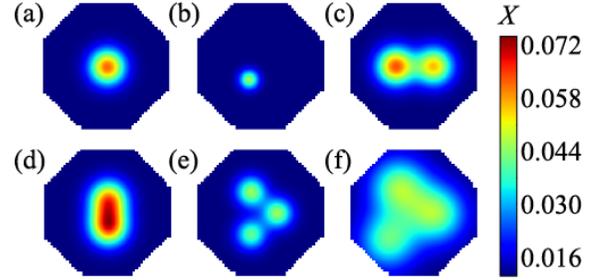

Fig. 7. True $H_2O$ concentration distributions with (a, b) one inhomogeneity, (c, d) two inhomogeneities, and (e, f) three inhomogeneities.

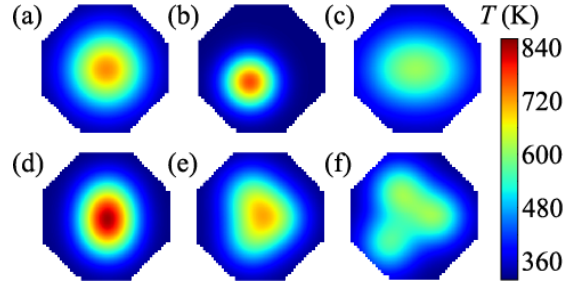

Fig. 8. True temperature distributions with (a, b) one inhomogeneity, (c, d) two inhomogeneities, and (e, f) three inhomogeneities.

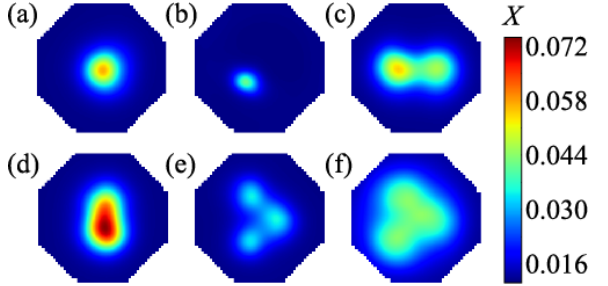

Fig. 9. Reconstructed $H_2O$ concentration distributions under an SNR of 35 dB for the corresponding phantoms shown in Fig. 7.

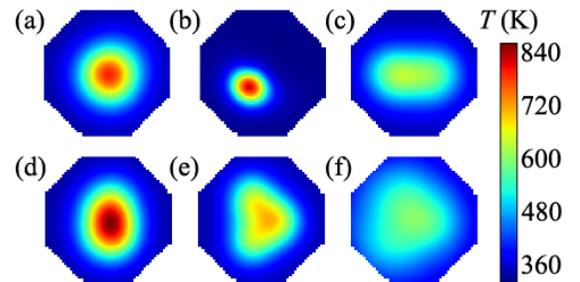

Fig. 10. Reconstructed temperature distributions under an SNR of 35 dB for the corresponding phantoms shown in Fig. 8.

*C. Test Results*

The established CSTNet was trained for three times with different randomness, yielding an ensemble of three different sets of trainable weights, i.e. $\Psi = \{\psi_1, \psi_2, \psi_3\}$. Image error (IE) of $H_2O$ concentration and temperature imaging at different signal-to-noise ratios (SNRs) is computed by

$$\text{IE}_{\text{conc.}} = \frac{1}{H}\sum_{h=1}^{H}\frac{\left\|X_h - \widehat{X_h}\right\|_2}{\left\|\widehat{X_h}\right\|_2} \tag{20}$$

and

$$\text{IE}_{\text{temp.}} = \frac{1}{H}\sum_{h=1}^{H}\frac{\left\|T_h - \widehat{T_h}\right\|_2}{\left\|T_h\right\|_2}, \tag{21}$$

where $h$ and $H$ denote the index and total number of test examples, respectively. $X_h$ ($T_h$) and $\widehat{X_h}$ ($\widehat{T_h}$) are the $h$-th true and reconstructed $H_2O$ concentration (temperature) distributions, respectively. The two images, noted by tuple $(\widehat{T_h}, \widehat{X_h})$, are simultaneously reconstructed by

$$\left(\widehat{T_h}, \widehat{X_h}\right) = \frac{1}{Z}\sum_{\psi \in \Psi} \pi_\psi\left(\left(A_{v_1}\right)_h^{test}, \left(A_{v_2}\right)_h^{test}\right), \tag{22}$$

where constant $Z$ denotes the size of the ensemble $\Psi$, and $\pi_\psi(\cdot)$ the end-to-end function of CSTNet given weights $\psi$. $\left(A_{v_1}\right)_h^{test}$ and $\left(A_{v_2}\right)_h^{test}$ are the $h$-th example from the test set.

As shown in Fig. 6, both $\text{IE}_{\text{conc.}}$ and $\text{IE}_{\text{temp.}}$ decrease as SNR decreases. With a practical SNR in real applications at approx. 35 dB, $\text{IE}_{\text{conc.}}$ and $\text{IE}_{\text{temp.}}$ are 0.1198 and 0.0714, respectively, denoting accurate retrieval of the true images. Furthermore, both $\text{IE}_{\text{conc.}}$ and $\text{IE}_{\text{temp.}}$ decrease for less than 10% when the SNR varies from 20 dB to 45 dB, indicating the proposed CSTNet is strongly robust to the measurement noise.

Six representative results with one, two, and three inhomogeneities are selected. As shown in Fig. 7 and Fig. 8, phantoms (a) and (b) include single inhomogeneity with different sizes and locations. Phantoms (c) to (f) have more inhomogeneities with different orientations, mutual distances, and sizes. As shown in Fig. 9 and Fig. 10, the reconstructions of these phantoms indicate the trained CSTNet can clearly distinguish the number of inhomogeneities, precisely locate the inhomogeneities, and accurately retrieve the true images. The proposed CSTNet contributes to supremely good quality of the tomographic images with no artefacts. Using Compute Unified Device Architecture (CUDA), we achieved simultaneous imaging with an average frame-per-second (fps) of 3,134, providing great potential for real-time multi-parameter imaging in industrial applications.

## V. EXPERIMENTS

In this section, laboratory experiments were carried out to further validate the proposed CSTNet model. The CST sensor was built in the optical layout depicted in Fig. 4. More details of the optics, e.g. lasers and detectors, electronics, e.g. data acquisition and signal processing system, and the parameter settings in wavelength modulation spectroscopy were described in our paper [36].

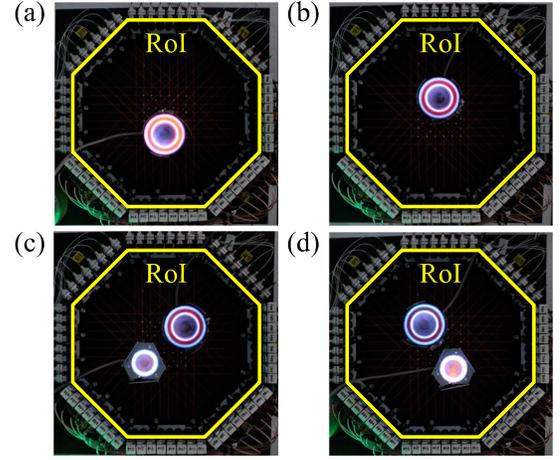

Fig. 11. Four reactive flow fields generated in the experiments with (a, b) a single flame and (c, d) two flames with different sizes.

As shown in Fig. 11, four cases with different distributions of $H_2O$ concentration and temperature were demonstrated in the experiments. The first two cases, shown in Figs. 11 (a) and (b), contain a single flame located at the lower centre and upper centre of the RoI, respectively. To consider more complex phantoms, two flames with different sizes and locations were generated in the other two cases shown in Figs. 11 (c) and (d).

The tomographic images of $H_2O$ concentration and temperature for the four cases are shown in Fig. 12 and Fig. 13, respectively. Retrieved peak values of the inhomogeneities in the tomographic images are listed in TABLE II. For the single-flame cases, locations of the flames in the tomographic images reconstructed by CSTNet agree well with original ones. As same flame is used in both cases, the similar retrieved peak values indicate good consistency between the reconstruction and the truth. For the dual-flame cases, the reconstructions not only precisely localise the two inhomogeneities, but also reveal their relative sizes. Last but not least, artefacts are significantly limited in all cases, demonstrating the proposed CSTNet is strongly robust for image reconstruction even with severely limited number of laser beams. The results are very promising for industry-oriented CST, mostly implemented in harsh environments, with low-complexity optical sensors.

It is worthwhile noting the proposed CSTNet shows advantages in the following three aspects:

1) Hybrid *a priori* information attributed to CST measurement is introduced by including the smoothness and the centrosymmetry, i.e. CST measurements of identical distributions at centrosymmetric positions in the RoI possess identical patterns. A shared feature extractor is established to directly learn these representations.
2) Fed by the above shared feature extractor, a dual-branch architecture with crosstalk decoders is built for simultaneous imaging of species concentration and temperature. The embedded crosstalk decoders learn their natural correlation, enabling more convincing and efficient image reconstruction in practical combustion processes.

3) An industry-orientated and low-complexity CST sensor containing only 32 laser beams is used, for the first time, to experimentally validate the proposed CSTNet. Aided by Graphic Processing Units (GPUs) and CUDA, CSTNet completes high-fidelity imaging within several milliseconds. The very short inference time enables online CST and further stimulates the industrial implementation of the proposed CSTNet on embedded devices, such as Field Programmable Gate Arrays (FPGAs) [37], [38].

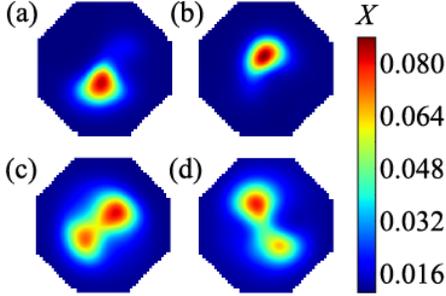

Fig. 12. Reconstructed $H_2O$ concentration distributions for the four cases in Fig. 11.

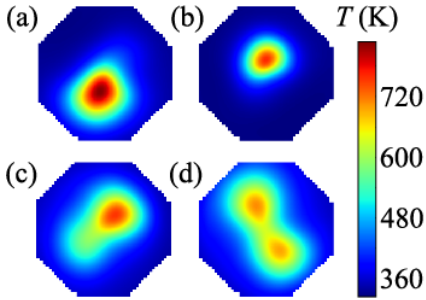

Fig. 13. Reconstructed temperature distributions for the four cases in Fig. 11.

TABLE II. Retrieved peak values for $H_2O$ concentration and temperature imaging. conc.: $H_2O$ concentration, temp.: temperature, L.: the larger flame, S.: the smaller flame.

|   | Peak conc. | Peak temp. (K) |   | Peak conc. | | Peak temp. (K) | |
|---|---|---|---|---|---|---|---|
|   |   |   |   | L. | S. | L. | S. |
| (a) | 0.084 | 828 | (c) | 0.082 | 0.073 | 755 | 600 |
| (b) | 0.088 | 761 | (d) | 0.080 | 0.065 | 708 | 695 |

## VI. CONCLUSION

In this paper, we developed a novel convolutional network named as CSTNet for simultaneous imaging of the distributions of species concentration and temperature in reactive flows. The inherently physical characteristics of the CST are learnt by a shared feature extractor, which incorporates the hybrid *a priori* information of smoothness and centrosymmetry. To simultaneously reconstruct the distributions of species concentration and temperature, a dual-branch architecture with crosstalk decoders is designed in CSTNet. The embedded crosstalk decoders take into account the natural correlation between species concentration and temperature, enabling more reasonable and efficient retrievals in practical combustion processes. The proposed CSTNet was both analytically and experimentally proven to be successful in high-fidelity imaging of $H_2O$ concentration and temperature images, using two $H_2O$ transitions and a tomographic sensor with 32 laser beams.

The performance of CSTNet was firstly evaluated using the simulated test set. Given a measurement SNR of 35 dB, CSTNet can accurately reconstruct various distributions of $H_2O$ concentration and temperature. For a wide range of SNR, numerical results indicate that CSTNet maintains excellent robustness against measurement noise. In the lab-scale experiments, image reconstruction using CSTNet achieves good agreement with the known locations of the original flames. The artefacts in the tomographic images are significantly eliminated, denoting strong resistance to the measurement noise in practical applications. Benefiting from GPU acceleration, the proposed CSTNet can simultaneously reconstruct images of $H_2O$ concentration and temperature distributions with 3,134 fps, exhibiting great potential for online CST towards real-time process control.

To the best of our knowledge, this is the first experimental application of deep learning to CST using an optical sensor with severely limited number of laser beams. In practice, the low-complexity optical sensor is overwhelmingly preferred in order to maintain the integrity of the industrial reactors and chambers. In our future work, we will deploy CSTNet on embedded devices such as FPGAs for the sake of more cost-efficient computation in industrial applications.